\documentclass[12pt, draftclsnofoot, onecolumn]{IEEEtran}

\usepackage{amsmath,amsfonts,amssymb,mathtools,dsfont,multicol,multirow,hhline,diagbox}
\usepackage{graphicx,subfigure}
\usepackage{algorithmic,algorithm,threeparttable,color}
\usepackage{amsthm,epstopdf}
\usepackage[noadjust]{cite}

\newtheorem{theorem}{Theorem}
\newtheorem{lemma}{Lemma}
\newtheorem{prop}{Proposition}

\newtheorem{corollary}{Corollary}

\ifCLASSINFOpdf
\else
\fi

\begin{document}
\title{A Unified Framework for the Tractable Analysis of  Multi-Antenna Wireless Networks}

\author{Xianghao~Yu,~\IEEEmembership{Student Member,~IEEE},
		Chang~Li,~\IEEEmembership{Member,~IEEE},
        Jun~Zhang,~\IEEEmembership{Senior Member,~IEEE},
	 Martin~Haenggi,~\IEEEmembership{Fellow,~IEEE},
        and~Khaled~B.~Letaief,~\IEEEmembership{Fellow,~IEEE}
\thanks{This work was presented in part at IEEE International Conference on Communications (ICC), Paris, France, May 2017 \cite{7997054}.}
\thanks{X. Yu, J. Zhang, and K. B. Letaief are with the Department of ECE, the Hong Kong University of Science and Technology (HKUST), Hong Kong (e-mail: \{xyuam, eejzhang, eekhaled\}@ust.hk).
C. Li is with the National Institute of Standards and Technology (NIST), Gaithersburg, MD 20899 USA (e-mail: chang.li@nist.gov), and this work was done when he was with HKUST.
M. Haenggi is with the Department of Electrical Engineering, University of
Notre Dame, Notre Dame, IN 46556 USA (e-mail: mhaenggi@nd.edu).

This work was supported by the Hong Kong Research Grants Council (No. 16210216) and the U.S.~NSF (grant CCF 1525904).
}
}

\maketitle

\begin{abstract}
	Densifying networks and deploying more antennas at each access point are two principal ways to boost the capacity of wireless networks.
	However, the complicated distributions of the signal power and the accumulated interference power, largely induced by various space-time processing techniques, make it highly challenging to quantitatively characterize the performance of  multi-antenna networks.
	In this paper, using tools from stochastic geometry, a unified framework is developed for the analysis of such networks. 
	The major results are two innovative representations of the coverage probability, which make the analysis of  multi-antenna networks almost as tractable as the single-antenna case. One is expressed as an $\ell_1$-induced norm of a Toeplitz matrix, and the other is given in a finite sum form.
	With a compact representation, the former incorporates many existing analytical results on single- and multi-antenna networks as special cases, and leads to tractable expressions for evaluating the coverage probability in both ad hoc and cellular networks. 
	While the latter is more complicated for numerical evaluation, it helps analytically gain key design insights. 
	In particular, it helps prove that the coverage probability of ad hoc networks is a monotonically decreasing convex function of the transmitter density and that there exists a peak value of the coverage improvement when increasing the number of transmit antennas.
	On the other hand, in multi-antenna cellular networks, it is shown that the coverage probability is independent of the transmitter density and that the outage probability decreases exponentially as the number of transmit antennas increases.
\end{abstract}

\IEEEpeerreviewmaketitle

\section{Introduction}
\subsection{Motivation}
To accommodate the ever-increasing mobile data traffic, there is a tremendous demand in boosting the capacity of wireless networks. One promising way is to exploit the spatial domain resources by deploying more antennas at transceivers, especially at the base station (BS), e.g., via the recently proposed massive MIMO technique \cite{6375940}.
Another effective way is via network densification \cite{bhushan2014network}, which can significantly improve the area spectral efficiency (ASE).
In this way,  multi-antenna networks form an important enabler for next-generation wireless networking \cite{andrews2014will}. Thus, it is of significant practical importance to understand the performance of such complicated networks.
While simulations can demonstrate many key features of  multi-antenna networks, mathematical analysis is needed to help expose their salient properties and provide effective mechanisms for comparing different design approaches without building and running system-level simulations.
However, the analysis of  multi-antenna networks is a highly challenging task, which may hinder their wide applicability.

To model the densely deployed transceivers, a random network model based on Poisson point processes (PPPs) has been extensively adopted to capture the irregularity and randomness of transmitter locations \cite{haenggi2012stochastic,baccelli2010stochastic}. With the help of stochastic geometry, this model turns out to be tractable, and the resulting aggregate interference can be analytically characterized \cite{6042301,6287527,6171997,haenggi2009interference}. Particularly, in single-antenna networks with Rayleigh fading channels, a large number of tractable results for various performance metrics have been derived based on the PPP model \cite{baccelli2010stochastic}.
When it comes to multi-antenna networks,  difficulties arise due to more complicated signal and interference distributions. These distributions are determined by two factors, namely the channel fading distribution and the adopted multi-antenna transmission techniques, which lead to a variety of highly challenging mathematical models to analyze.
While significant efforts have been made, so far there is no systematic methodology to analyze  multi-antenna networks. To fill this gap, in this paper we propose a unified analytical framework for such networks, which is almost as tractable as that for single-antenna networks. More importantly, it leads to important system design insights for different network models.

\subsection{Related Works}
Adopting the PPP model for the analysis of cellular networks was first advocated in \cite{baccelli2010stochastic}, which derived tractable results for the coverage and ergodic rate analysis mainly assuming Rayleigh fading channels, i.e., exponential distributed
channel power gains. It disclosed that the coverage probability is critically determined by the Laplace transform of the aggregate interference. While this study inspired many research works on  network analysis and design, e.g., \cite{6171997,6047548}, the main results are only applicable to single-antenna networks.

Considering the important role of MIMO techniques, there have been many works trying to derive tractable results for multi-antenna wireless networks \cite{5288965,4712724,7880697,5668921,6157054,6596082,6932503,7478073,7156167,6881662,6587514,5673756,6205593}. 
The first-order Taylor expansion has been used to provide approximations for the coverage probability in two-tier heterogeneous networks (HetNets) \cite{5288965} and the transmission capacity of ad hoc networks \cite{4712724}.
In \cite{7880697}, the fading in the interferers' channels was ignored to obtain a tight lower bound on the ergodic spectral efficiency, instead of the usual approach of calculating $\mathbb{E}[\log(1+\mathrm{SINR})]$, which leads to a looser upper bound since it implicitly assumes that the serving BS knows all the channel states.
Upper bounds have also been derived for signal-to-interference ratio (SIR) outage probabilities in mobile ad hoc networks \cite{5668921,6157054} and cellular networks \cite{6596082,6932503}. In particular, different inequalities, e.g., Markov's inequality \cite{5668921}, Chebyshev's inequality \cite{6157054}, the union bound \cite{6596082},
and an upper bound for the cumulative distribution function of gamma random variables \cite{6932503}, 
have been adopted to make the analysis tractable. Nevertheless, with \emph{approximations} or \emph{upper bounds}, the  results obtained may not accurately reflect the behavior of multi-antenna networks in all operating regimes. On the other hand, exact analytical evaluations were investigated in \cite{7478073,7156167,6881662,6587514,5673756,6205593}. The downlink spectral efficiency and rate coverage probability were derived in \cite{7478073} and \cite{7156167}, respectively, with improper integrals that are inefficient for numerical evaluation. In addition, closed-form expressions for coverage probabilities were obtained in \cite{6881662,6587514,5673756,6205593}. However, these results are stated in complicated forms (nested sums) with special functions, such as the complementary incomplete beta function \cite{6881662,6587514}, and Stirling numbers of the first and second kind \cite{5673756,6205593}. These existing results, either approximate or with very bulky expressions, are not able to yield insights for network design and optimization.

Recently, some promising results were reported in \cite{6775036,7038201,7412737,7997054,7913628}, where closed-form expressions were derived for various performance metrics in multi-antenna HetNets and further used to solve practical system design problems. In particular, different multi-antenna transmission techniques over different fading channels were considered, e.g., maximum ratio transmission (MRT) and zero-forcing (ZF) over Rayleigh fading channels \cite{6775036,7038201}, space-division multiple access (SDMA) in multi-tier HetNets over Rayleigh fading channels \cite{7412737}, jamming in physical layer security-aware networks \cite{7997054}, and analog beamforming in millimeter wave (mm-wave) networks with Nakagami fading channels \cite{7913628}.
These results were derived for specific scenarios. In contrast, in this paper we will develop a unified framework to analyze  multi-antenna networks based on which key system insights are then revealed.

\subsection{Contributions}
In this paper, we analyze multi-antenna networks with a random spatial network model, where transmitters are modeled as a homogeneous PPP \cite{haenggi2012stochastic}. 
The analytical results in this paper are listed in Table \ref{tab2}, and the main contributions are summarized as follows.
\begin{table}[t]
	\centering
	\caption{Summary of the analytical results}\label{tab2}
	\begin{tabular}{|cc|c|c|}\hline
		&       & \textbf{Cellular networks} & \textbf{Ad hoc networks} \\\hhline{|=|=|=|=|}
		\textbf{Coverage probability} & \multicolumn{1}{|c|}{General channel power gain} & Proposition \ref{prop2}  & Proposition \ref{prop1} \\\cline{2-4}
		\textbf{expression}& \multicolumn{1}{|c|}{Gamma distributed channel power gain} & Corollary \ref{coro2}  & Corollary \ref{coro1} \\\hline
		\multirow{2}[0]{*}{\textbf{Unique properties}} & \multicolumn{1}{|c|}{Transmitter density} & Invariant  & Corollary \ref{coro3} \\\cline{2-4}
		& \multicolumn{1}{|c|}{Antenna size} & Proposition \ref{prop3} \& Corollary \ref{coro4}  & Propositions \ref{prop5} \& \ref{prop4} \\\hline
	\end{tabular}%
\end{table}%
\begin{itemize}
	\item We develop a unified analytical framework for  multi-antenna networks, which is applicable to networks where the signal power gain is gamma distributed while the interferers' power gains have arbitrary distributions. 
	In the proposed framework, the recursive relations between the $n$-th derivatives of the Laplace transform are exploited, based on which two novel representations of the coverage probability are derived in Theorems \ref{the1} and \ref{th1}, namely an \emph{$\ell_1$-Toeplitz matrix representation} and a \emph{finite sum representation}. 
	We demonstrate that this new framework makes the analysis of  multi-antenna networks almost as tractable as the single-antenna case \cite{6042301}. More importantly, many analytical techniques developed for conventional single-antenna networks can be easily transplanted to the general multi-antenna setting. 
	
	\item With the $\ell_1$-Toeplitz matrix representation, a general coverage expression is given for cellular networks and ad hoc networks, as presented in Propositions \ref{prop2} and \ref{prop1}. Compared with existing works, these analytical expressions are not only expressed in more compact forms but also provide an exact characterization for  multi-antenna networks.
	
	\item With the finite sum representation, the impacts of the antenna size and network density are investigated in both cellular and ad hoc networks. 
	It is analytically shown in Corollary \ref{coro3} that, when the transmitter density increases, the SIR coverage probability of ad hoc networks is a monotone decreasing convex function of the transmitter density.
	In addition, the outage probability of cellular networks  decreases exponentially when increasing the number of antennas, as shown in Proposition \ref{prop3}. In contrast, Propositions \ref{prop5} and \ref{prop4} demonstrate that there may exist a peak value for the coverage improvement in ad hoc networks. 
\end{itemize}

\subsection{Organizations and Notations}
The remainder of this paper is organized as follows. Section \ref{II} presents the system model along with the unified analytical framework. The proposed framework is then applied to cellular and ad hoc networks in Section \ref{III}. In Section \ref{IV}, we reveal the impact of the antenna size and network density on the coverage probability. Finally, Section \ref{V} concludes the paper.

In this paper, matrices are denoted by bold-face upper-case letters, and the $M\times M$ identity matrix is represented as $\mathbf{I}_M$. The $\ell_1$-induced matrix norm is defined as $\left\Vert\mathbf{A}\right\Vert_1=\max_{1\le j\le n}\sum_{i=1}^{m}|a_{ij}|$ for $\mathbf{A}\in\mathbb{R}^{m\times n}$. The expectation is denoted as $\mathbb{E}[X]$, and the probability that an event $A$ happens is denoted as $\mathbb{P}(A)$. The gamma function, lower incomplete gamma function, and upper incomplete gamma function are denoted as $\Gamma(x)$, $\gamma(s,x)$, and $\Gamma(s,x)$, respectively. The $n$-th derivative of the function $f(x)$ is denoted as $f^{(n)}(x)$. The falling factorial of a number $x$ is symbolized as $(x)_{n}$. The generalized hypergeometric function is denoted as ${}_pF_q\left(\{a_i\}_{i=1}^p;\{b_i\}_{i=1}^q;z\right)$ \cite[Sec. 9.14]{zwillinger2014table}.

\section{A Unified Analytical Framework}\label{II}
\subsection{System Model}
Consider the downlink transmission of a multi-antenna wireless network. 
We focus on the performance analysis of the typical receiver at the origin, whose signal-to-interference-plus-noise ratio (SINR) is given by\footnote{The expression in \eqref{SINR1} can describe the SINR for each transmitted data stream in spatial multiplexing \cite{6205593}, or the SINR in multi-tier networks given that the typical receiver is associated with a certain tier \cite{5288965,6171996,7412737}.}
\begin{equation}\label{SINR1}
\mathrm{SINR}=\dfrac{g_{x_0} r_0^{-\alpha}}{\sigma_\mathrm{n}^2 + \sum_{x\in\Phi^\prime}g_{x} \|x\|^{-\alpha}}.
\end{equation}
To clarify the generality of the proposed framework, the notations and assumptions underlying \eqref{SINR1} are explained below.
\begin{itemize}
	\item $r_0$: the distance from the typical receiver to its associated transmitter located at $x_0$, i.e., $r_0\triangleq\Vert x_0\Vert$. It can be either a deterministic value, or a random variable with the probability density function (pdf) denoted as $f_{r_0}(r)$. 
	The choice of $x_0$ depends on the adopted cell association strategy which is typically based on the long-term average received power, e.g., dipole association in ad hoc networks \cite{4712724}, the nearest-transmitter association in cellular networks \cite{6042301}, and biased association in HetNets \cite{7412737}.
	
	\item $\sigma_\mathrm{n}^2$: the normalized noise power, where the normalization is to keep expression \eqref{SINR1} clean. For example, it is given as $\sigma_\mathrm{n}^2=\frac{U\sigma^2}{P}$ when assuming equal power allocation to serve $U$ receivers, where $P$ is the transmit power.

	\item $g_{x_0}$: the channel power gain for the desired signal  from the associated transmitter located at $x_0$.
	 In particular, assuming there are in total $D$ ($D\ge1$) data streams that are simultaneously transmitted via spatial multiplexing, a general form of the channel power gain for the $d$-th ($d\in\{1,\dots,D\}$) data stream with linear MIMO transmission/reception techniques is given by \cite{6205593}
	\begin{equation}\label{eq2}
	g_{x_0}=\frac{\left|\left[\mathbf{W}_{x_0}\mathbf{H}_{x_0}\mathbf{F}_{x_0}\right]_d\right|^2}{\left[\mathbf{W}_{x_0}\mathbf{W}_{x_0}^H\right]_d},
	\end{equation} 
	where $[\mathbf{A}]_d$ represents the $d$-th diagonal element of matrix $\mathbf{A}$, and $\mathbf{H}_{x_0}$ denotes the channel matrix from the associated transmitter to the typical receiver. In addition, $\mathbf{F}_{x_0}$ and $\mathbf{W}_{x_0}$ are the beamforming and combining matrices, which are determined by the adopted MIMO transmission scheme.
	
	Different channel distributions and MIMO techniques lead to different distributions for $g_{x_0}$. In this paper, a general type of distribution is assumed for $g_{x_0}$, as specified below.
	
	\emph{\textbf{Assumption 1}}: The channel power gain $g_{x_0}$ for the desired signal is gamma distributed, i.e., $g_{x_0}\sim\mathrm{Gamma}(M,\theta)$, where $M$ and $\theta$ are the shape and scale parameters of the gamma distribution. 
	
	Table \ref{tab1} lists some commonly-used multi-antenna transmission techniques and channel fading distributions and the corresponding distributions of $g_{x_0}$. It is shown that the gamma distribution is typically encountered in the analysis of multi-antenna networks. Moreover, our proposed framework can be applied to more general distributions, as will be discussed later in Remark 3. To keep the presentation clean, the analytical results throughout the paper are based on the gamma distribution.
	\begin{table}[t]
		\centering
		\caption{Typical multi-antenna transmission techniques and corresponding signal power gain distributions}
		\begin{tabular}{|l|c|c|c|}
			\hline
			& {\textbf{Multi-antenna transmission}} &\textbf{Channel}& {\textbf{Signal power gain}} \\
			&\textbf{technique ($\mathbf{F}_{x_0}$/$\mathbf{W}_{x_0}$)}&\textbf{fading ($\mathbf{H}_{x_0}$)}&\textbf{ ($g_{x_0}$) distribution}\\\hhline{|=|=|=|=|}
			\textbf{Throughput and Energy}  &\multirow{2}[0]{*}{MRT}   & \multirow{2}[0]{*}{Rayleigh} & \multirow{2}[0]{*}{$\mathrm{Gamma}(N_\mathrm{t},1)$} \\
			\textbf{Efficiency Analysis} \cite{6775036}     &       &&  \\\hline
			\textbf{Interference Coordination} \cite{7038201}& Partial ZF beamforming&Rayleigh&$\mathrm{Gamma}(\max(N_\mathrm{t}-N_{x_0},1),1)$\\\hline
			\textbf{SIMO Ad Hoc Networks} \cite{5668921}&Partial ZF combining &Rayleigh&$\mathrm{Gamma}(N_\mathrm{r}-N_{x_0},1)$\\\hline
			\textbf{Spatial Multiplexing}	& {Maximum ratio combining} &\multirow{2}[0]{*}{Rayleigh} & \multirow{2}[0]{*}{$\mathrm{Gamma}(N_\mathrm{r},1)$} \\
			\textbf{in Ad Hoc Networks} \cite{5673756}    & (MRC)&      &  \\\hline		
			\textbf{Multi-tier Multiuser}	& \multirow{2}[0]{*}{SDMA} &\multirow{2}[0]{*}{Rayleigh} & \multirow{2}[0]{*}{$\mathrm{Gamma}(N_{\mathrm{t}}-U+1,1)^*$} \\
			\textbf{MIMO HetNets} \cite{7412737}    & &      &  \\\hline
			\textbf{Physical Layer Security}     &Jamming \&&\multirow{2}[0]{*}{Rayleigh}& \multirow{2}[0]{*}{$\mathrm{Gamma}(D,1)$} \\
			\textbf{Aware Networks} \cite{7997054}    & ZF beamforming &     &  \\\hline
			\textbf{Millimeter-wave Networks} \cite{7913628}	& Analog beamforming &Nakagami&$\mathrm{Gamma}(N_\mathrm{t},1/N_\mathrm{t})$\\\hline
		\end{tabular}%
		\label{tab1}%
		\begin{tablenotes}
			\item * The parameters are for each tier in HetNets.
			\item The numbers of antennas at the transmitter and receiver sides are denoted as $N_\mathrm{t}$ and $N_\mathrm{r}$, respectively, and $U$ denotes the number of served users in SDMA systems. In addition, $N_{x_0}$ represents the number of transmitters that the typical receiver requests to perform interference canceling. Please refer to the corresponding references for more details.
		\end{tablenotes}
	\end{table}%
	\item $\Phi^\prime$: the set of interfering transmitters. This set can be a union of different sets consisting of $J$ ($J\ge 1$) types of interferers, i.e., $\Phi^\prime=\cup_{j=1}^J\Phi_j^\prime$. In particular, the interferers belonging to the $j$-th type are distributed according to a PPP $\Phi_j^\prime$ conditioned on $b(o,l_j(r_0))$ to be empty, where $b(z, R)$ is a disk centered at $z$ with radius $R$, $l_j(r_0)$ is the minimum distance between the typical receiver and the transmitter of the $j$-th type, and the functions $l_j(\cdot)$ are determined by the cell association strategy. 
	This model not only reflects the general multi-tier HetNet setting but also applies when the interferers have different channel power gain distributions in a single-tier network.
	\item $g_{x}$: the interferer's power gain from the interfering transmitter located at $x$. A concrete expression for $g_{x}$ similar to \eqref{eq2} can be derived \cite[eq. (2)]{6205593}. In the proposed framework, we assume that  $\big\{(g_{x})_{x\in\Phi^\prime_j}\big\}_{j=1}^J$ are $J$ families of non-negative random variables that are independent and identically distributed according to arbitrary distributions for which the $\frac{2}{\alpha}$-th moments exist.
	\item $\alpha$: the path loss exponent. In the proposed framework, $\alpha$ can assume any value larger than 2. $\alpha=4$  is the typical value used in many previous works to simplify the analytical results.
\end{itemize}


\subsection{Analytical Framework for  Multi-Antenna Networks}
There are various performance metrics for wireless networks, e.g, outage probability,  ASE, average throughput, and energy efficiency. Note that one fundamental task in characterizing these metrics is to calculate the SINR distribution \cite{6775036,7038201,7412737,haenggi2012stochastic}. In this paper, we focus on deriving the complementary cumulative distribution function (ccdf) of the SINR, also called  the \emph{coverage probability}, which is defined as\footnote{For a $K$-tier HetNet,  
	the coverage probability $p_{\mathrm{c},k}(\tau)$ given that the typical receiver is associated with the $k$-th tier can be calculated by \eqref{coveragedef}, and the overall coverage probability is then given by $\sum_{k=1}^KA_kp_{\mathrm{c},k}(\tau)$, where $A_k$ is the probability that the typical receiver is associated with the $k$-th tier. }
\begin{equation}\label{coveragedef}
	p_\mathrm{c}(\tau)\triangleq\mathbb{P}(\mathrm{SINR}>\tau),
\end{equation}
where $\tau$ denotes the SINR threshold. Its complement is the \emph{outage probability}, defined as $p_\mathrm{o}(\tau)\triangleq1-p_\mathrm{c}(\tau)$.

In this section, we provide a unified analytical framework for multi-antenna wireless networks.
First, according to the SINR expression \eqref{SINR1}, the coverage probability defined in \eqref{coveragedef} can be written as
\begin{equation}
	p_\mathrm{c}(\tau)
	=\mathbb{P}\left[g_{x_0}>\tau r_0^\alpha\left(\sigma_\mathrm{n}^2 +I\right)\right],\label{coverageprob}
\end{equation}
where $I\triangleq\sum_{x\in\Phi^\prime}g_x \|x\|^{-\alpha}$.
One main difficulty of the analysis comes from the gamma distributed random variable $g_{x_0}$. Different from existing works that adopted approximations \cite{5288965,4712724} or upper bounds \cite{5668921,6157054,6932503},  we derive a compact and exact expression for the probability \eqref{coverageprob}. According to the ccdf of the gamma distribution,  \eqref{coverageprob} is firstly rewritten as
\begin{align}
	p_\mathrm{c}(\tau)&=\mathbb{E}_{r_0}\left\{\sum_{n=0}^{M-1}\frac{(\tau r_0^\alpha/\theta)^n}{n!}\mathbb{E}_I\left[\left.(\sigma_\mathrm{n}^2+I)^ne^{-\frac{\tau r_0^\alpha}{\theta} (\sigma_\mathrm{n}^2+I)}\right|r_0\right]\right\}\nonumber\\
	&=\mathbb{E}_{r_0}\left[\sum_{n=0}^{M-1}\frac{(-s)^n}{n!}\mathcal{L}^{(n)}(s)\right],\label{eq4}
\end{align}
where $s\triangleq\tau r_0^\alpha/\theta$, and $\mathcal{L}(s)\triangleq e^{-s\sigma_\mathrm{n}^2}\mathbb{E}_I\left[\left.e^{-sI}\right|r_0\right]$ is the Laplace transform of noise and interference conditioned on the distance $r_0$.
According to the probability generating functional (PGFL) of PPP \cite{haenggi2012stochastic}, 
the conditional Laplace transform $\mathcal{L}(s)$ can be expressed in a general exponential form as
\begin{align}
\mathcal{L}(s)&=\exp\left\{-s\sigma_\mathrm{n}^2-2\pi\sum_{j=1}^J\lambda_j\int_{l_j(r_0)}^\infty\left(1-\mathbb{E}_{g_j}[\exp(-sg_jv^{-\alpha})]\right)v\mathrm{d}v\right\}\label{Ls3}\\
&\triangleq\exp\{\eta(s)\}\nonumber,
\end{align}
 where $\lambda_j$ is the density\footnote{\color{black}For network models incorporating load awareness \cite{6775036,6918448}, the activation of transmitters can be reflected in the density $\lambda_j$.} of $\Phi_j^\prime$, the interferers' power gain is denoted as $g_{j}$ that is identically distributed as all the $\left(g_{x}\right)_{x\in\Phi_j^\prime}$. Here we use $\eta(s)$ to simplify the notation, which is called the \emph{log-Laplace transform}.
 
{\color{black}\emph{Remark 1:} 
	The proposed framework does not depend on the form of the log-Laplace transform, and it can be readily extended to other network models, for example, where the transmitters are spatially distributed according to other point processes \cite{8187697}, or the multi-slope path loss model is considered \cite{7061455}. One can first determine the log-Laplace transform $\eta(s)$ according to the network model and then the analytical framework can be applied similarly. In addition, as established in \cite{7322270}, the SIR coverage probability of cellular non-Poisson models is well approximated by $p_\mathrm{c}(\tau/G)$, where $p_\mathrm{c}(\tau)$ is the coverage probability of the cellular Poisson model and $G$ is a gain factor that depends on the geometry of the non-Poisson model. Hence, the results in this paper permit a simple approximation of the coverage probabilities for any stationary and ergodic point process model.}

As shown in \eqref{eq4}, the main task in deriving the coverage probability in multi-antenna networks is to calculate the $n$-th derivatives of the Laplace transform $\mathcal{L}(s)$. In single-antenna networks with Rayleigh fading channels, this operation is not needed, as the signal power gain is exponentially distributed.
While there exist some approaches to calculate the $n$-th derivative of a general exponential function, e.g., via Fa\`a di Bruno's formula \cite{roman1980formula} or Bell polynomials \cite{zwillinger2014table}, a direct computation of the derivatives leads to unwieldy expressions \cite{6881662,5673756,6205593,6205588}, which cannot be efficiently evaluated and fail to reveal system insights. 

Instead of working with the Laplace transform directly, we analyze the log-Laplace transform $\eta(s)$. As we will see, this approach results in tractable results for the coverage probability.
First, the recursive relation between the derivatives of the Laplace transform is revealed in the following lemma.
\begin{lemma}\label{lem1}
	Defining $p_n=\frac{(-s)^n}{n!}\mathcal{L}^{(n)}(s)$, there exist recursive relations between $\{p_n\}_{n=0}^{\infty}$, given by
	\begin{equation}\label{recursive}
		p_n=\sum_{i=0}^{n-1}\frac{n-i}{n}t_{n-i}p_i,
	\end{equation}
	where
	\begin{equation}\label{nd}
	t_k=\frac{(-s)^k}{k!}\eta^{(k)}(s).
	\end{equation}
\end{lemma}
\begin{IEEEproof}
	First, it is obvious that $p_0=\mathcal{L}(s)=e^{\eta(s)}$ and $\mathcal{L}^{(1)}(s)=\eta^{(1)}(s)\mathcal{L}(s)$. According to the formula of Leibniz for the $n$-th derivative of the product of two functions \cite{roman1980formula}, we have
	\begin{equation}
	\mathcal{L}^{(n)}(s)=\frac{\mathrm{d}^{n-1}}{\mathrm{d}s}\mathcal{L}^{(1)}(s)=\sum_{i=0}^{n-1}{{n-1}\choose i} \eta^{(n-i)}(s)\mathcal{L}^{(i)}(s),
	\end{equation}
	followed by
	\begin{equation}
	\frac{(-s)^n}{n!}\mathcal{L}^{(n)}(s)=\sum_{i=0}^{n-1}\frac{n-i}{n}\frac{(-s)^{(n-i)}}{(n-i)!}\eta^{(n-i)}(s)\frac{(-s)^i}{i!}\mathcal{L}^{(i)}(s),
	\end{equation}
	which completes the proof by applying the definition that $p_n=\frac{(-s)^n}{n!}\mathcal{L}^{(n)}(s)$.
\end{IEEEproof}

According to the recursive relations in \eqref{recursive} and the fact that $p_0=\mathcal{L}(s)$, the only factors we need to calculate to obtain $\{p_n\}_{n=1}^{M-1}$ are the coefficients $\{t_k\}_{k=0}^{M-1}$, which are related to the  derivatives of $\eta(s)$. So the main task is shifted from calculating the derivatives of $\mathcal{L}(s)$ to deriving those of $\eta(s)$.
As shown in extensive existing works \cite{7997054,5288965,4712724,5668921,6157054,7478073,7156167,6881662,6587514,5673756,6205593,6775036,7038201,7412737},  obtaining a closed-form solution for $\eta^{(n)}(s)$ is generally much easier than for $\mathcal{L}^{(n)}(s)$, which will be further demonstrated in this paper.
Following Lemma \ref{lem1}, a finite sum representation of the coverage probability is given in the following theorem.
\begin{theorem}\label{the1}
	\emph{\textbf{(Finite Sum Representation of the Coverage Probability)}} The coverage probability \eqref{coverageprob} is given by
	\begin{equation}\label{finitesum}
	p_\mathrm{c}(\tau)=\mathbb{E}_{r_0}\left[\sum_{n=0}^{M-1}p_n\right].
	\end{equation}
	 where $\{p_n\}_{n=0}^{M-1}$ are given in Lemma \ref{lem1}.
\end{theorem}
\begin{IEEEproof}
	The result follows from \eqref{eq4} and the definition of $p_n$ in Lemma \ref{lem1}.
\end{IEEEproof}



\emph{Remark 2:} The main merit of this representation is that it leads to valuable system insights. For example, the impact of the shape parameter $M$ in the gamma distribution, which is typically related to the antenna size, is clearly illustrated by this finite sum representation. We define  $\bar{p}_n\triangleq\mathbb{E}_{r_0}[p_n]$ to simplify the presentation. In particular, as shown in Table \ref{tab1} and the references therein, the interferers' power gains $g_{x,j}$ are typically independent of $M$ with various MIMO transmission techniques, and so are $\{\bar{p}_n\}_{n=0}^{M-1}$. When $M$ increases, e.g., from $M$ to $M+\Delta$, the number of terms in the sum increases, and the variation of the coverage probability is directly related to the coefficients $\{\bar{p}_n\}_{n=M}^{M+\Delta-1}$.
This property will be leveraged to reveal the impact of the antenna size in Section \ref{IV}.

From both \eqref{eq4} and \eqref{finitesum}, it is apparent that the main challenge in evaluating the coverage probability is to derive a tractable expression for $\{p_n\}_{n=0}^{M-1}$. 
With Theorem \ref{the1}, we need to calculate $\{p_n\}_{n=0}^{M-1}$ in a recursive manner, which is still tedious.
Next, we derive more explicit expressions for $\{p_n\}_{n=0}^{M-1}$, assuming that we have obtained $\{t_k\}_{k=0}^{M-1}$. To this end, we define the two power series
\begin{equation}\label{eq40}
T(z)\triangleq\sum_{n=0}^\infty t_nz^n,\quad 
P(z)\triangleq\sum_{n=0}^\infty p_nz^n.
\end{equation}

\begin{lemma}\label{lem2}
	The power series $P(z)$ is related to $T(z)$ as
	\begin{equation}
	P(z)=e^{T(z)}.\label{exp}
	\end{equation}
\end{lemma}
\begin{IEEEproof}
It is straightforward to show that $T^{(1)}(z)=\sum_{n=0}^{\infty}(n+1)t_{n+1}z^{n}$ and $P^{(1)}(z)=\sum_{n=0}^{\infty}np_nz^{n-1}$. We then have the following equality
\begin{equation}
T^{(1)}(z)P(z)=\sum_{n=0}^\infty\sum_{i=0}^{n-1}(n-i)t_{n-i}p_iz^{n-1}.
\end{equation}
Combined with \eqref{recursive}, we obtain the differential equation
\begin{equation}
P^{(1)}(z)=T^{(1)}(z)P(z),
\end{equation}
whose solution is given by \eqref{exp}.
\end{IEEEproof}

Based on Lemma \ref{lem2}, an explicit expression for the coverage probability is given in the following theorem, which is more tractable than the result in Theorem \ref{the1}.
\begin{theorem}\emph{\textbf{($\ell_1$-Toeplitz Matrix Representation of the Coverage Probability)}}\label{th1} The coverage probability \eqref{coverageprob} is given by
	\begin{equation}\label{frameexpr}
	p_\mathrm{c}(\tau)=
	\mathbb{E}_{r_0}\left[\left\Vert e^{\mathbf{T}_M}\right\Vert_1\right],
	\end{equation}
	where $\mathbf{T}_M$ is the following $M\times M$ lower triangular Toeplitz matrix
	\begin{equation}
	\mathbf{T}_M=\left[{\begin{IEEEeqnarraybox*}[][c]{,c/c/c/c/c,}
		t_0&{}&{}&{}&{}\\
		t_1&t_0&{}&{}&{}\\
		t_2&t_1&t_0&{}&{}\\
		\vdots &{}&{}& \ddots &{}\\
		t_{M-1}&\cdots&t_2& t_1 &t_0
		\end{IEEEeqnarraybox*}} \right],\label{topmatrix}
	\end{equation}
	and its non-zero entries are determined by \eqref{nd}.
\end{theorem}
\begin{IEEEproof}
According to \eqref{finitesum}, \eqref{eq40}, and \eqref{exp}, the coverage probability is given by
\begin{equation}\label{eq19}
p_\mathrm{c}(\tau)=\mathbb{E}_{r_0}\left[\sum_{n=0}^{M-1}p_n\right]=\mathbb{E}_{r_0}\left[\sum_{n=0}^{M-1}\frac{1}{n!}\left.{P^{(n)}(z)}\right|_{z=0}\right]=\mathbb{E}_{r_0}\left[\sum_{n=0}^{M-1}\frac{1}{n!}\frac{\mathrm{d}^n}{\mathrm{d}z^n}\left.{e^{T(z)}}\right|_{z=0}\right].
\end{equation}
In the last expression of \eqref{eq19}, the $n$-th term in the sum is determined by the $n$-th coefficient of the power series $e^{T(z)}$. From \cite[pp. 14]{henrici1974applied}, the first $M$ coefficients of the power series $e^{T(z)}$ form the first column of the matrix exponential $e^{\mathbf{T}_M}$, whose exponent is given in \eqref{topmatrix}. The sum of these coefficients can be written as an $\ell_1$-induced matrix norm as in \eqref{frameexpr}.
\end{IEEEproof}

Compared with the approximations in \cite{4712724,5288965,5668921,6157054,6932503} and complicated expressions in \cite{7478073,7156167,6881662,6587514,5673756,6205593}, the $\ell_1$-Toeplitz matrix representation in \eqref{frameexpr} provides a more compact form for the coverage probability. More importantly, it enables us to leverage various powerful tools from linear algebra, especially some nice properties of lower triangular Toeplitz matrices, to provide insightful design guidelines for network optimization. Such properties can be found in \cite{6775036}, where they were used for small-cell networks. 



\emph{Remark 3:} A more general form of the pdf of  $g_{x_0}$ that may be encountered in multi-antenna wireless networks is given in \cite[eq. (10)]{6205593} as
\begin{equation}\label{eq21}
f_{g_{x_0}}(u)=\sum_{p\in\mathcal{P}}e^{-\phi_pu}\sum_{q\in\mathcal{Q}}\varphi_{p,q}u^q,
\end{equation}
where $\mathcal{P,Q}\subset\mathbb{N}_0$ and $\phi_p,\varphi_{p,q}\in\mathbb{R}$ are model parameters. In addition, various special cases of this pdf with different MIMO transmission techniques, e.g., transmit
antenna selection with ZF receivers and open-loop spatial multiplexing with ZF receivers, are specified in \cite[Table I]{6205593}. 
According to \eqref{eq21}, the ccdf of $g_{x_0}$ is given by
\begin{equation}
F^c_{g_{x_0}}(u)=1-\sum_{p\in\mathcal{P}}\sum_{q\in\mathcal{Q}}\frac{\varphi_{p,q}q!}{\phi_p^{q+1}}+
\sum_{p\in\mathcal{P}}\sum_{q\in\mathcal{Q}}\frac{\varphi_{p,q}q!}{\phi_p^{q+1}}\sum_{k=0}^qe^{-\phi_qu}\frac{(\phi_pu)^k}{k!}.
\end{equation}
The proposed framework is also applicable to this general form of pdf, resulting in the coverage probability
\begin{equation}\label{eq23}
p_\mathrm{c}(\tau)=1-\sum_{p\in\mathcal{P}}\sum_{q\in\mathcal{Q}}\frac{\varphi_{p,q}q!}{\phi_p^{q+1}}+
\sum_{p\in\mathcal{P}}\sum_{q\in\mathcal{Q}}\frac{\varphi_{p,q}q!}{\phi_p^{q+1}}
\mathbb{E}_{r_0}\big[\big\Vert e^{\mathbf{T}_{q+1}^{(p)}}\big\Vert_1\big],
\end{equation}
where $\mathbf{T}_{q+1}^{(p)}$ in the $p$-th term in the sum denotes a $(q+1)\times (q+1)$ lower triangular Toeplitz matrix similar to \eqref{topmatrix}. Furthermore, the non-zero entries in $\mathbf{T}_{q+1}^{(p)}$ are given by
\begin{equation}
t_{p,k}=\frac{(-s_p)^k}{k!}\eta^{(k)}(s_p), \quad 0\le k\le q,
\end{equation}
where $s_p=\tau r_0^\alpha\phi_p$. Note that the result in Theorem \ref{the1} corresponding to the gamma distribution $\mathrm{Gamma}(M,\theta)$ is a special case of \eqref{eq23} with $\mathcal{P}=\{0\}$, $\mathcal{Q}=\{M-1\}$, $\phi_0=\frac{1}{\theta}$, and $\varphi_{0,M-1}=\frac{1}{\theta^M\Gamma(M)}$.
In this paper, to keep the presentation concise and easy to follow, we use the gamma distribution to present the main context, but all the results in this paper are applicable for the general pdf in \eqref{eq21}.

\emph{Remark 4:} When applying Theorem \ref{th1} to specific multi-antenna networks, the only parameters to be determined are the non-zero entries $\{t_n\}_{n=0}^{M-1}$ in the matrix $\mathbf{T}_M$, and the main steps for applying the proposed framework are summarized as Methodology 1.
As mentioned before, Theorem \ref{th1} is a generalization of our previous results in \cite{6775036,7038201,7412737}. 

Although an additional expectation over $r_0$ is needed when the distance between the typical receiver and its associated transmitter is a random variable, e.g., in cellular networks, in the next section we will show that closed-form expressions are available via the proposed framework. As listed in Table \ref{tab3}, in the remainder of this paper, Theorems \ref{the1} and \ref{th1} will be utilized to analyze multi-antenna networks in specific settings.

\floatname{algorithm}{Methodology}
\begin{algorithm}[t]
	\label{alternating}
	\caption{Main Steps to Apply the Proposed Framework}
	\begin{algorithmic}[1]
		\STATE 
		Derive the conditional Laplace transform $\mathcal{L}(s)$ according to \eqref{Ls3} for the given distributions of $\{g_j\}_{j=1}^J$ and the specific point processes for the interfering transmitters $\{\Phi_j^\prime\}_{j=1}^J$;
		\STATE 
		Calculate the $n$-th ($1\le n\le M-1$) derivatives of $\eta(s)$ to populate the entries $\{t_n\}_{n=0}^{M-1}$ in the matrix $\mathbf{T}_M$ according to \eqref{nd};
		\STATE
		Express the coverage probability $p_\mathrm{c}(\tau)$ with Theorem \ref{th1}.
	\end{algorithmic}
\end{algorithm}
\begin{table}[t]
	\centering
	\caption{The use of Theorems \ref{the1} and \ref{th1} in Sections \ref{III} and \ref{IV}.}
	\begin{tabular}{|l|c|c|c|}\hline
		\multirow{2}[0]{*}{} & \textbf{Corollaries \ref{coro2} and \ref{coro1} }      &\textbf{Corollary \ref{coro4}} & \multirow{2}[0]{*}{\textbf{Corollary \ref{coro3}}} \\
		&  \textbf{Propositions \ref{prop2}, \ref{prop1}, and \ref{prop6}}     &  \textbf{Propositions \ref{prop3}, \ref{prop5}, and \ref{prop4}}     &  \\\hline
		\textbf{Theorem \ref{the1} }   &       &      \checkmark &\checkmark  \\\hline
		\textbf{Theorem \ref{th1}}    & \checkmark      &       &\checkmark  \\\hline
	\end{tabular}%
	\label{tab3}%
\end{table}%

\subsection{Single-Antenna vs. Multi-Antenna Networks}\label{II-C}
Here we show that our proposed framework incorporates the single-antenna network as a special case. Assuming Rayleigh fading, the signal power gain is exponentially distributed in the single-antenna case, i.e., $M=\theta=1$. In this way, expression \eqref{frameexpr} in Theorem \ref{th1} (or \eqref{finitesum} in Theorem 1) simplifies to
\begin{equation}
	p_\mathrm{c}(\tau)=\int_0^\infty f_{r_0}(r)\mathcal{L}(s)\mathrm{d}r,
\end{equation}
which is exactly the  classic result in \cite[Prop. 7.3.1]{baccelli2010stochastic}. Note that, for single-antenna networks, the main task to derive the coverage probability is to derive the conditional Laplace transform $\mathcal{L}(s)$. It has been shown in \cite{baccelli2010stochastic} that, under various assumptions for the interferers' power gain, $\mathcal{L}(s)$ (equivalently $\eta(s)$) can be derived in closed form. This in turn makes it possible to express the coverage probability in a closed form.

%

When it comes to multi-antenna networks, Theorem \ref{th1} is compatible with any forms of $\eta(s)$.
Furthermore, with the gamma distributed signal power gain, according to Remark 4, the only additional task compared with single-antenna networks is to calculate $M-1$ derivatives of $\eta(s)$, which does not introduce much computational complexity and thus preserves the tractability. This means that many manipulation tricks and steps developed for single-antenna networks, e.g., derivation techniques listed in \cite[Sec. III]{6524460}, can be adopted to the multi-antenna case. The tractability and effectiveness of the proposed framework will be illustrated in the next section by developing new analytical results for general ad hoc and cellular networks.

\section{Coverage Analysis for  Multi-Antenna Networks}\label{III}
In this section, based on the general framework, we specify the analytical results for cellular and ad hoc networks. By leveraging the $\ell_1$-Toeplitz matrix representation in Theorem \ref{th1}, tractable expressions for the coverage probability are provided.
Single-tier networks are considered to keep the presentation neat, but the derivation can be easily extended to general HetNets by calculating the Laplace transform according to \eqref{Ls3}. Furthermore, since  wireless networks are interference-limited, we focus on the SIR distribution instead of SINR.

\subsection{Coverage Analysis in Cellular Networks}
In the cellular network model considered in this paper, the typical user is associated with the nearest BS. Thus, the pdf of the distance $r_0$ between the typical user and the serving BS is given by \cite{1512427}
\begin{equation}
f_{r_0}(r)=2\pi\lambda re^{-\pi \lambda r^2},
\end{equation}
and the SIR is expressed as
\begin{equation}\label{cSINR}
\mathrm{SIR}=\dfrac{g_{x_0} r_0^{-\alpha}}{\sum_{x\in\Phi\backslash\{x_0\}}g_x \|x\|^{-\alpha}}.
\end{equation}
Since the nearest BS is part of the PPP $\Phi$ consisting of all the transmitters, the set of interfering BSs $\Phi^\prime=\Phi\backslash\{x_0\}$ forms a
PPP on $\mathbb{R}^2\backslash b(0,r_0)$ conditioned on $x_0\in\Phi$. Recall that in Assumption 1 we assumed that $g_{x_0}$ is a gamma distributed random variable, i.e., $g_{x_0}\sim\mathrm{Gamma}(M,\theta)$. We define $\delta\triangleq\frac{2}{\alpha}$ and let $g$ be a random variable identically distributed as all the $(g_x)_{x\in\Phi^\prime}$, which are two notations that shall be frequently used in this paper. 
\begin{prop}\label{prop2}
	When the locations of BSs are modeled as a PPP, and the nearest-BS association is adopted in the cellular network, the SIR coverage probability is given by
	\begin{equation}\label{eq25}
	p_\mathrm{c}(\tau)=\left\Vert\mathbf{C}_M^{-1}\right\Vert_1,
	\end{equation}
	with the non-zero entries in the lower triangular Toeplitz matrix $\mathbf{C}_M$ as
	\begin{equation}\label{eq26}
	c_n=\frac{\delta}{\delta-n}\frac{\left({\tau}/{\theta}\right)^n}{n!}\mathbb{E}_g\left[g^n{}_1F_1\left(n-\delta;n+1-\delta;-\frac{\tau}{\theta}g\right)\right],\quad 0\le n\le M-1.
	\end{equation}
\end{prop}
\begin{IEEEproof}
A detailed proof is provided here to illustrate the main steps in applying the proposed framework for the coverage analysis. The proofs for the remaining results follow similar steps and are therefore  diverted to the appendix.

	We first simplify the expression in Theorem \ref{th1} under the cellular network model. According to the two-step approach of applying Theorem \ref{th1} as presented in Remark 4, first we calculate the log-Laplace transform as 
	\begin{equation}\label{eq27}
	\begin{split}
	\eta(s)&=-2\pi\lambda\int_{r_0}^\infty{\left(1-\mathbb{E}_{g}[\exp(-sgv^{-\alpha})]\right)}v\mathrm{d}v\\
	&\overset{(a)}{=}\pi\lambda r_0^2+\pi\lambda \delta s^\delta\mathbb{E}_g\left[g^\delta\gamma(-\delta,sr^{-\alpha}_0g)\right]\\
	&\overset{(b)}{=}\pi\lambda r_0^2-\pi\lambda r_0^2\mathbb{E}_g\left[{}_1F_1\left(-\delta;1-\delta;-sr_0^{-\alpha}g\right)\right],
	\end{split}
	\end{equation}
	where (a) can be derived from \cite[eq. (4)]{6042301} by changing variables $v^{-\alpha}\to y$, and step (b) applies the identity $\gamma(s,x)\equiv\frac{x^s}{s}{}_1F_1(s,s+1,-x)$ \cite[Sec. 6.45]{zwillinger2014table}.
	Then, by utilizing the derivatives
	\begin{equation}\label{eq32}
	\frac{\mathrm{d}^n}{\mathrm{d}z^n}{}_1F_1\left(a;b;z\right)=\frac{\prod_{p=0}^{n-1}(a+p)}{\prod_{p=0}^{n-1}(b+p)}{}_1F_1\left(a+n;b+n;z\right),
	\end{equation}  
	the non-zero entries in $\mathbf{T}_M$ in \eqref{topmatrix} are determined by \eqref{nd}, i.e.,
	\begin{equation}\label{eq28}
	\begin{split}
	t_n&=\frac{(-s)^n}{n!}\eta^{(n)}(s)\\
	&=
	-\pi\lambda r_0^2\frac{\delta}{\delta-n}\frac{\left({\tau}/{\theta}\right)^n}{n!}\left\{\mathbb{E}_g\left[g^n{}_1F_1\left(n-\delta;n+1-\delta;-\frac{\tau}{\theta}g\right)\right]-\mathds{1}(n=0)\right\}\\
	&=
	-\pi\lambda r_0^2\left[c_n-\mathds{1}(n=0)\right],
	\end{split}
	\end{equation}
	where $\{c_n\}_{n=0}^{M-1}$ are given in \eqref{eq26} and  $\mathds{1}(\cdot)$ denotes the indicator function. The coverage probability is evaluated following \eqref{frameexpr} as
	 \begin{equation}
	 p_\mathrm{c}(\tau)=\int_0^\infty 2\pi\lambda re^{-\pi \lambda r^2}\left\Vert e^{\mathbf{T}_M}\right\Vert_1\mathrm{d}r.
	 \end{equation}
	 This formula can be further simplified into a closed form by defining a power series similar to \eqref{eq40}, i.e., $C(z)=\sum_{n=0}^\infty c_nz^n$. According to \eqref{eq28}, we have
	 \begin{equation}
	 T(z)=\sum_{n=0}^\infty t_nz^n=\pi\lambda r_0^2\left(1-c_0-\sum_{n=1}^\infty c_nz^n\right)=\pi\lambda r_0^2\left[1-C(z)\right].
	 \end{equation}
	 To help the derivation, another power series $\bar{P}(z)=\sum_{n=0}^{\infty}\bar{p}_nz^n$ is defined as
	 \begin{equation}\label{eq33}
	 \bar{P}(z)\triangleq\mathbb{E}_{r_0}\left[P(z)\right].
	 \end{equation}
	Hence, the power series $\bar{P}(z)$ is written as
	\begin{equation}\label{eq31}
	\begin{split}
		\bar{P}(z)&= \mathbb{E}_{r_0}\left[P(z)\right]\overset{(c)}{=}\mathbb{E}_{r_0}\left[e^{T(z)}\right]
		=\int_0^\infty 2\pi\lambda re^{-\pi\lambda r^2}e^{T(z)}\mathrm{d}r\\
		&=\int_0^\infty 2\pi\lambda re^{-\pi\lambda C(z)r^2}\mathrm{d}r=\frac{1}{C(z)},
	\end{split}
	\end{equation}
	where (c) is due to Lemma \ref{lem2}.
	Applying Theorem \ref{the1} and \eqref{eq31}, we have
	\begin{equation}\label{eq36}
	p_\mathrm{c}(\tau)=\sum_{n=0}^{M-1}\bar{p}_n=\sum_{n=0}^{M-1}\frac{1}{n!}\left.{\bar P^{(n)}(z)}\right|_{z=0}=\sum_{n=0}^{M-1}\frac{1}{n!}\frac{\mathrm{d}^n}{\mathrm{d}z^n}\left.{\frac{1}{C(z)}}\right|_{z=0}.
	\end{equation}
	Similar to what we exploited in \eqref{eq19}, from \cite[pp. 14]{henrici1974applied}, the first $M$ coefficients of the power series $\frac{1}{C(z)}$ form the first column of the matrix inversion $\mathbf{C}_M^{-1}$, and their sum is the  $\ell_1$-induced matrix norm of $\mathbf{C}^{-1}_M$ as given in \eqref{eq25}.
\end{IEEEproof}

\emph{Remark 5:} This result expresses the coverage probability of cellular networks in a very compact form, where only an inverse of a lower triangular Toeplitz matrix is needed. There exist many fast algorithms to calculate this inverse \cite{commenges1984fast}, which makes \eqref{eq25} more efficient than existing analytical results, e.g., \cite{6881662,6587514}.
In addition, the class of models for which this result applies is also more general.

While general in the interferers' power gain, Proposition \ref{prop2} loses some tractability due to the expectation over $g$ when calculation $\{c_n\}_{n=0}^{M-1}$. The following corollary presents a more tractable expression for a specific distribution for the interferers' power gain, i.e., $g\sim\mathrm{Gamma}(\kappa,\beta)$. Note that this is a commonly encountered distribution for the interferers' power gain in multi-antenna networks, as previously shown in \cite{4712724,5288965,5668921,6157054,7478073,7156167,6881662,6587514,5673756,6205593}. 
\begin{corollary}\label{coro2}
	Under Assumption 1, when the interferers' power gain is gamma distributed as $g\sim\mathrm{Gamma}(\kappa,\beta)$, the SIR coverage probability of cellular networks is given by
	\begin{equation}\label{equation36}
	p_\mathrm{c}(\tau)=\left\Vert\mathbf{C}_M^{-1}\right\Vert_1,
	\end{equation}
	with the non-zero entries in $\mathbf{C}_M$ as
	\begin{equation}
	c_n=\frac{\Gamma(\kappa+n)}{\Gamma(\kappa)\Gamma(n+1)}\frac{\delta}{\delta-n  }\left(\frac{\tau\beta}{\theta}\right)^n{}_2F_1\left(n+\kappa,n-\delta;n+1-\delta;-\frac{\tau \beta}{\theta}\right),\quad0\le n\le M-1.
	\end{equation}
\end{corollary}
\begin{IEEEproof}
	See Appendix \ref{appB}.
\end{IEEEproof}

\emph{Remark 6:} This result is a generalization of our previous works \cite{6775036,7038201,7412737} where the parameters $\kappa$ and $\beta$ are specified for different network settings. The non-zero entries $c_0$ and $\{c_n\}_{n=1}^{M-1}$ were obtained by  two different expressions in \cite{6775036,7038201}, and they are now unified in Corollary \ref{coro2}. 

\begin{figure}[t]
	\centering\includegraphics[height=5.9cm]{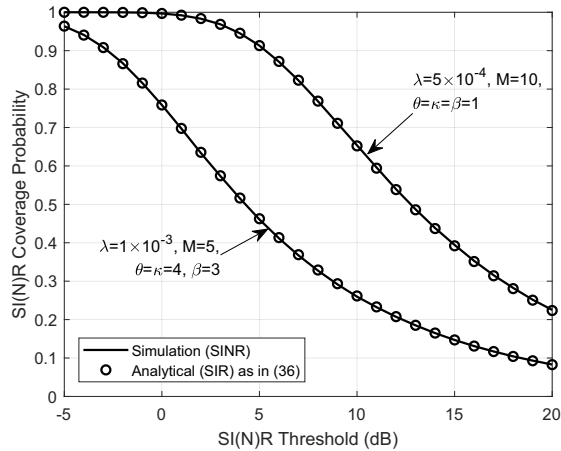}
	\caption{The SI(N)R coverage probability of cellular networks when $\alpha=4$, and $\sigma_\mathrm{n}^2=-97.5$ dBm.}\label{fig2}
\end{figure}

Fig. \ref{fig2} plots the SIR coverage probability of cellular networks with \eqref{equation36}. In addition, our analytical results are shown to be accurate even if noise is included, which verifies the interference dominance assumption.
\subsection{Coverage Analysis in Ad Hoc Networks}
In ad hoc networks, a dipole model is adopted. Specifically, a dipolar pair is added with its receiver at
the origin, which becomes the typical pair under expectation over the point process, and therefore $\Phi^\prime=\Phi$. The communication distance $r_0$ between the typical receiver and its associated transmitter is assumed to be fixed as the dipole distance \cite{haenggi2012stochastic}, and the nearest interferer can be arbitrarily close to the typical receiver. This means that, in ad hoc networks, there is no need to calculate the integral over $r_0$ in \eqref{frameexpr}. The resulting SIR is 
\begin{equation}\label{adSINR}
	\mathrm{SIR}=\dfrac{g_{x_0} r_0^{-\alpha}}{ \sum_{x\in\Phi}g_x \|x\|^{-\alpha}},
\end{equation}
where the signal power gain $g_{x_0}$ is gamma distributed per Assumption 1.
Correspondingly, the coverage probability in ad hoc networks is given by the following proposition.
\begin{prop}\label{prop1}
	Under Assumption 1, the SIR coverage probability of ad hoc networks is given by\footnote{The matrix $\mathbf{A}_M$ has the same expression as $\mathbf{T}_M$ in \eqref{frameexpr}. The change of notation here is mainly to distinguish the results in ad hoc networks from those under general network settings.}
	\begin{equation}\label{eq35}
	p_\mathrm{c}(\tau)=\left\Vert e^{\mathbf{A}_M}\right\Vert_1,
	\end{equation}
	where $\mathbf{A}_M$ is the lower triangular Toeplitz matrix with the non-zero entries as
	\begin{equation}\label{an1}
		a_n=-\frac{(-1)^n}{n!}(\delta)_n\pi\lambda r_0^2\Gamma(1-\delta)\left(\frac{\tau}{\theta}\right)^\delta\mathbb{E}_g\left[g^\delta\right],\quad0\le n\le M-1.
	\end{equation}
\end{prop}
\begin{IEEEproof}%
See Appendix \ref{appA}.
\end{IEEEproof}

\emph{Remark 7:} In the ad hoc network model, even if the noise is included, it is still feasible to derive a closed-form expression for the coverage probability. Specifically, the log-Laplace transform is given by
\begin{equation}\label{eq37}
\eta(s)=-s\sigma_\mathrm{n}^2-\pi\lambda\Gamma(1-\delta)s^\delta\mathbb{E}_g\left[g^\delta\right].
\end{equation}
Hence, the non-zero entries $\{a_n\}_{n=0}^{M-1}$ in \eqref{an1} are 
\begin{equation}\label{an3}
a_n=\frac{(-s)^n}{n!}\eta^{(n)}(s)=\frac{(-1)^n}{n!}
\left\{-\mathds{1}(n\le1)\frac{\tau r_0^\alpha}{\theta}\sigma_\mathrm{n}^2-
\pi\lambda r_0^2\Gamma(1-\delta)(\delta)_n\left(\frac{\tau}{\theta}\right)^\delta\mathbb{E}_g\left[g^\delta\right]
\right\}.
\end{equation}

\emph{Remark 8:} Proposition 2 expresses the coverage probability of ad hoc networks by an  $\ell_1$-induced norm of a matrix exponential. 
In particular, once the distribution of the interferers' power gain $g$ is given, the non-zero entries $\{a_n\}_{n=0}^{M-1}$ in the lower triangular matrix $\mathbf{A}_M$ can be obtained according to \eqref{an1} or \eqref{an3}. Finally, a matrix exponential is the only operation needed in the calculation. Efficient techniques exist for computing the matrix exponential of lower triangular Toeplitz matrices \cite{kressner2016fast}.

Similar to cellular networks, next we present a special case with closed-form expressions where the interferers' power gain is gamma distributed as $g\sim\mathrm{Gamma}(\kappa,\beta)$. 
\begin{corollary}\label{coro1}
	Under Assumption 1, when the  interferers' power gain is gamma distributed as $g\sim\mathrm{Gamma}(\kappa,\beta)$, the SINR coverage probability of ad hoc networks is given by
	\begin{equation}
	p_\mathrm{c}(\tau)=\left\Vert e^{\mathbf{A}_M}\right\Vert_1,
	\end{equation}
	with the non-zero entries in $\mathbf{A}_M$ as
	\begin{equation}\label{an2}
a_n=\frac{(-1)^n}{n!}\left\{-\mathds{1}(n\le1)\frac{\tau r_0^\alpha}{\theta}\sigma_\mathrm{n}^2-\pi\lambda r_0^2\left(\frac{\tau\beta}{\theta}\right)^\delta\frac{\Gamma(\delta+\kappa)\Gamma(1-\delta)\Gamma(1+\delta)}{\Gamma(\kappa)\Gamma(\delta+1-n)}\right\},\quad0\le n\le M-1.
	\end{equation}
\end{corollary}
\begin{IEEEproof}
	The result follows by inserting $\mathbb{E}\left[g^\delta\right]=\beta^\delta\Gamma(\delta+\kappa)/\Gamma(\kappa)$ in \eqref{an3}.
\end{IEEEproof}


\section{Unique Properties in Cellular and Ad Hoc Networks}\label{IV}
In the previous section, the $\ell_1$-Toeplitz matrix representation in Theorem \ref{th1} has been applied to derive tractable expressions for the coverage in cellular and ad hoc networks.
In this section, the finite sum representation in Theorem \ref{the1}, assisted by Theorem \ref{th1}, is applied to reveal unique properties in both types of networks. We investigate the effects of the transmitter density and the transmitter antenna size on the coverage probability as examples. 

\subsection{The Effect of Network Density}
For cellular networks modeled by stationary point processes, scaling the plane by any factor does not change the SIR if  nearest-BS association is adopted with the homogeneous path loss law. For instance, taking $c\Phi$ ($c>0$) instead of $\Phi$, which equivalently scales the network density by $c^{-2}$, does not affect the coverage probability. In other words, very generally, the coverage probability in cellular networks is invariant to the BS density $\lambda$. This SIR invariance property has been revealed in some specific settings, e.g., \cite{baccelli2010stochastic,6042301}.

On the other hand, in ad hoc networks, since the distance between the typical receiver and the associated transmitter is fixed, the coverage probability monotonically decreases when the transmitter density increases, as the densification implies more interferers per unit area. However, there is no existing works that quantified such effect, which is pursued in the following result.
\begin{corollary}\label{coro3}
	The SIR coverage probability \eqref{eq35} is a monotonically decreasing convex function of the transmitter density,
	and it can be rewritten as
	\begin{equation} \label{eq41}
	p_\mathrm{c}(\lambda)= e^{a_0^\prime\lambda}\sum_{n=0}^{M-1}\beta_n\lambda^n,
	\end{equation}
	where
	\begin{equation}
	\beta_n=
	\frac{\left\Vert \left(\mathbf{A}^\prime_M-a_0^\prime\mathbf{I}_M\right)^n\right\Vert_1}{n!},
	\end{equation}
	and
	\begin{equation}
	a_n^\prime=\frac{a_n}{\lambda}=-\frac{(-1)^n}{n!}(\delta)_n\pi r_0^2\Gamma(1-\delta)\left(\frac{\tau}{\theta}\right)^\delta\mathbb{E}_g\left[g^\delta\right].
	\end{equation}
	Correspondingly, the derivative of the coverage probability with respect to the transmitter density is given by
	\begin{equation}\label{eq49}
	\frac{\partial}{\partial \lambda}p_\mathrm{c}(\lambda)= e^{a_0^\prime\lambda}\left\{a_0^\prime\beta_{M-1}\lambda^{M-1}+\sum_{n=0}^{M-2}\left[a_0^\prime\beta_n+(n+1)\beta_{n+1}\right]\lambda^n\right\}.
	\end{equation}
\end{corollary}
\begin{IEEEproof}
	See Appendix \ref{appC}.
\end{IEEEproof}

\begin{figure}[t]
	\centering\includegraphics[height=5.9cm]{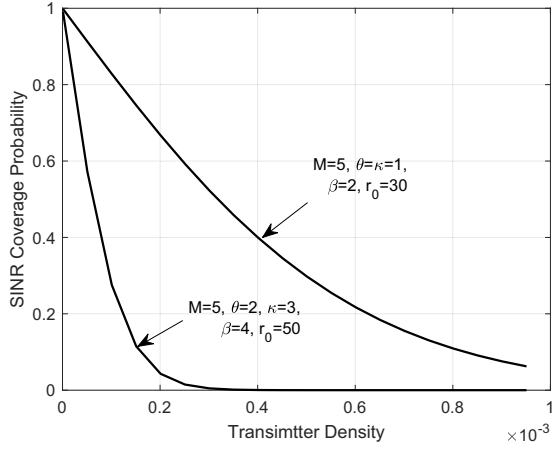}
	\caption{The impact of the transmitter density on the SIR coverage probability in ad hoc networks when $\alpha=4$ and $\tau=0$ dB, according to \eqref{eq41}.}\label{fig3}
\end{figure}

From Corollary \ref{coro3} we have $p_\mathrm{c}(\lambda)\to 1$ as $\lambda\to0$, which is independent of  all the other network parameters. Hence, for any coverage requirement $1-\epsilon$ at the typical receiver, there exists a maximum transmitter density $\lambda$ that can satisfy it regardless of the other network parameters, and this density can be numerically determined.
Furthermore, this result fully characterizes how the transmitter density affects the coverage probability, which is shown in Fig. \ref{fig3}. In particular, we prove that increasing the transmitter density  degrades the coverage probability in ad hoc networks, and the coverage probability is a product of an exponential function and a polynomial function of order $M-1$ of the transmitter density $\lambda$. 
For the special case of $M=1$, i.e., single-antenna networks with Rayleigh fading channel, the coverage probability reduces to an exponential one. In other words, the multi-antenna setting increases the coverage probability by the additional polynomial term. In addition, the derivative given in \eqref{eq49} reflects the sensitivity of the coverage probability with respect to the transmitter density.

\emph{Remark 9:} A related result on the impact of the dipole distance $r_0$ can be readily obtained from Corollary \ref{coro3}. Since $r_0^2$ and $\lambda$ are interchangeable in \eqref{eq41}, there exists a duality between $\lambda$ and $r_0^{-2}$, where the former affects the interference power while the latter only affects the signal power. The impact of the dipole distance $r_0$ is then given by
\begin{equation}
p_\mathrm{c}(r_0)= e^{\hat{a}_0 r_0^2}\sum_{n=0}^{M-1}\hat{\beta}_nr_0^{2n},
\end{equation}
where
$
\hat{\beta}_n=
\frac{\left\Vert \left(\mathbf{\hat{A}}_M-\hat{a}_0\mathbf{I}_M\right)^n\right\Vert_1}{n!}
$
and
$
\hat{a}_n={a_n}/{r_0^2}
$.
Similar to Corollary \ref{coro3}, it can be proved that the coverage probability is a monotonically decreasing convex function of the dipole distance.

\emph{Remark 10:} The monotonicity and convexity in Corollary \ref{coro3} are also applicable to the SINR coverage probability where the noise is also taken into consideration, and the proof can be found in Appendix \ref{appC}.

\emph{Remark 11:} As shown in Appendix \ref{appC}, Corollary \ref{coro3} is obtained based on the proposed analytical framework in Section \ref{II}. Particularly, its derivation is greatly simplified by the delicate tackling of the gamma distributed signal power, via the  representations derived. If the analytical results in existing works \cite{4712724,5673756,6587514} were used instead, we would not be able to explicitly disclose the impact of the transmitter density, which, from another perspective, confirms the advantages of the proposed analytical framework.

\subsection{The Effect of the Antenna Size}
As discussed in Section \ref{II-C}, our proposed framework generalizes our ability in analyzing the single-antenna network to the multi-antenna one. Hence, it is intriguing to apply it to investigate how multi-antenna techniques affect the coverage probability. In the following, we shall perform such an investigation by taking a MISO network with MRT beamforming as an example. In this case, the signal power gain $g_{x_0}$ is gamma distributed as $\mathrm{Gamma}(M,\theta)$, where $M$ is the number of transmit antennas.

\emph{Remark 12:} As shown in Table \ref{tab1}, the number of antennas is typically related to the shape parameter $M$ in the gamma distribution of the signal power gain $g_{x_0}$. 
Hence, the derivations and conclusions in the following are also applicable to other network parameters related to the shape parameter $M$, e.g., the user number $U_k$ in MIMO HetNets \cite{7412737}, the number of coordination requests $K_{x_0}$ in user-centric interference coordination \cite{7038201}, and the number of transmitted streams $N_{x_0}$ in physical layer security-aware networks \cite{7997054}.

We first present a general lemma that will be used in the following derivation. We define the \emph{coverage improvement} for the $n$-th antenna as the increment of the coverage probability when the antenna size is enlarged from $n-1$ to $n$.
\begin{lemma}\label{lem3}
	For both ad hoc and cellular networks, the coverage improvement due to the $M+1$-th antenna is
	\begin{equation}
	p_\mathrm{c}(M+1)-p_\mathrm{c}(M)=\bar p_M.
	\end{equation}
	For ad hoc networks, $\bar p_n=p_n$ while for cellular $\bar p_n=\mathbb{E}_{r_0}[p_n]$, with $\{{p}_n\}_{n=0}^\infty$ given in Lemma \ref{lem1}.
\end{lemma}
\begin{IEEEproof}
	The result follows directly from Theorem \ref{the1}.
\end{IEEEproof}
 Intuitively, enlarging the antenna size  increases both the information signal power as well as the interference power,  hence an explicit analysis is needed to reveal the overall effect. Based on Lemma \ref{lem3}, we have the following result.
 \begin{prop}\label{prop6}
 For both ad hoc and cellular networks, increasing the antenna size always improves the coverage probability, i.e., $\bar{p}_n>0$ for $n>0$.
 \end{prop}
\begin{IEEEproof}
	According to \eqref{frameexpr} and \eqref{eq62}, we have
	\begin{equation}
	p_\mathrm{c}=\mathbb{E}_{r_0}\left[\left\Vert e^{\mathbf{T}_M}\right\Vert_1\right]=\mathbb{E}_{r_0}\left[e^{t_0}
	\left(1+\sum_{n=1}^{M-1}\frac{\left\Vert\left(\mathbf{T}_M-t_0\mathbf{I}_M\right)^n\right\Vert_1}{n!}\right)\right].
	\end{equation}
	Hence, $\bar{p}_n$ can be rewritten as
	\begin{equation}
	\bar{p}_n=\mathbb{E}_{r_0}\left[e^{t_0}
	\frac{\left\Vert\left(\mathbf{T}_M-t_0\mathbf{I}_M\right)^n\right\Vert_1}{n!}\right].
	\end{equation}
	Similar to \eqref{eq61}, it can be proved that $t_0<0$ while $t_n>0$ for $n>0$. In this way, all the entries in the strict lower triangular matrix $\mathbf{T}_M-t_0\mathbf{I}_M$ are non-negative, and so are $\{\bar{p}_n\}_{n=0}^\infty$.
\end{IEEEproof}
Note that Proposition \ref{prop6} applies to very general network settings, as long as the signal power gain is gamma distributed, as stated in Assumption 1, and the assumption that the shape parameter $M$ is the only parameter related to the number of antennas. In the following, we apply this result to different network models.

\begin{prop}\label{prop3}
	Denoting the outage probability in multi-antenna cellular networks by $p_\mathrm{o}(M)$, we have
		\begin{equation}
		\underset{M\to\infty}{\lim}\frac{p_\mathrm{o}(M)}{p_\mathrm{o}(M+1)}=\underset{n\to\infty}{\lim}\frac{\bar{p}_n}{\bar p_{n+1}}=r_\mathrm{c}>1,
		\end{equation}
		where $r_\mathrm{c}$ is the radius of convergence of the power series $\bar{P}(z)$ in \eqref{eq31}, given by the solution to the equation
		\begin{equation}\label{equation47}
		\mathbb{E}_g\left[{}_1F_1\left(-\delta;1-\delta;\frac{(r_\mathrm{c}-1)\tau}{\theta}g\right)\right]=0.
		\end{equation}
\end{prop}  
\begin{IEEEproof}
	See Appendix \ref{appD}.
\end{IEEEproof}

A corollary of this result is given next when the interferers' power gain is gamma distributed as $g\sim\mathrm{Gamma}(\kappa,\beta)$.
\begin{corollary}\label{coro4}
	When the interferers' power gain in  multi-antenna cellular networks is gamma distributed, we have
	\begin{equation}
		\underset{M\to\infty}{\lim}\frac{p_\mathrm{o}(M)}{p_\mathrm{o}(M+1)}=r_\mathrm{c},
	\end{equation}
	where $r_\mathrm{c}\in\left(1,1+\frac{\theta}{\tau\beta}\right)$ is the solution to the equation
	\begin{equation}
	{}_2F_1\left(\kappa,-\delta;1-\delta,(r_\mathrm{c}-1)\frac{\tau \beta}{\theta}\right)=0.
	\end{equation}
\end{corollary}
\begin{IEEEproof}
	It is proved by plugging the pdf of $g$, i.e., $f_g(u)=\frac{u^{\kappa-1}e^{-\frac{u}{\beta}}}{\beta^\kappa\Gamma(\kappa)}$, into \eqref{equation47}, and the upper bound of $r_\mathrm{c}$ can be easily obtained from the radius of convergence of the Gaussian hypergeometric function.
\end{IEEEproof}

\emph{Remark 13:} Proposition \ref{prop3} indicates that, when $M$ is large, the coverage improvement of adding the $n$-th antenna is ${r_\mathrm{c}}$ times larger than that of adding the $(n+1)$-th antenna. Furthermore, the outage probability of cellular networks in the logarithmic scale decreases linearly in $M$ with slope $-\log_{10}{r_\mathrm{c}}$.

\begin{figure}[t]
	\centering\includegraphics[height=5.9cm]{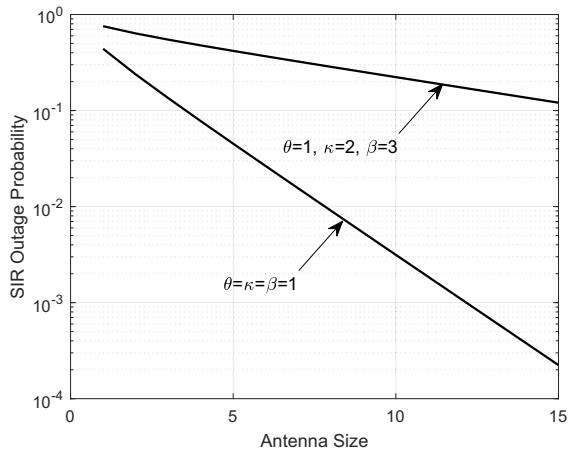}
	\caption{The SIR outage probability of cellular networks when $\tau=0$ dB and $\alpha=4$, according to \eqref{equation36}.}\label{fig4}
\end{figure}

Fig. \ref{fig4} shows the SIR outage probability of cellular networks versus the antenna size. While Proposition \ref{prop3} is an asymptotic result, it is quite accurate also when the number of antennas is small. In addition, as $r_\mathrm{c}$ is larger than 1 in Proposition \ref{prop3}, it demonstrates that increasing the antenna size definitely benefits the coverage probability, and it also shows that the coverage improvement $\bar{p}_n$ diminishes as the number of antennas grows large. However, this may not be the case in ad hoc networks, as shown next.

Analyzing the coverage improvement in ad hoc networks for general network settings is more challenging, so we start from the  special case $\alpha=4$, which is usually used in existing works \cite{6042301,6932503} for analytical tractability. Particularly, we focus on finding the antenna index, denoted by $n^\star+1$, that contributes the most significant coverage improvement in ad hoc networks.

\begin{prop}\label{prop5}
	When the path loss exponent $\alpha=4$, the SIR coverage improvement due to adding the $n+1$-th antenna in ad hoc multi-antenna networks monotonically decreases in the interval
	\begin{equation}
	n>\frac{\mu^2}{4}-1,
	\end{equation}
	where $\mu>0$ is given by
	\begin{equation}
	\mu=\pi\lambda r_0^2\Gamma(1-\delta)\left(\frac{\tau}{\theta}\right)^\delta\mathbb{E}_g\left[g^\delta\right].
	\end{equation}
\end{prop}
\begin{IEEEproof}
	See Appendix \ref{appF}.
\end{IEEEproof}

\emph{Remark 14:} Proposition \ref{prop5} indicates that the largest coverage improvement occurs when adding one of the first $\left\lceil\frac{\mu^2}{4}-1\right\rceil+1$ antennas, i.e., $1\le n^\star\le \left\lceil\frac{\mu^2}{4}-1\right\rceil$. Furthermore, the condition that the coverage improvement is always monotonically decreasing can be derived via Proposition \ref{prop5}, given by $\frac{\mu^2}{4}-1<0$, i.e., $\mu<2$.

\begin{figure}[t]
	\centering
	\subfigure[The increment of the SIR coverage probability of ad hoc networks when $\mathrm{SNR}=0$ dB, $\theta=\kappa=\beta=1$, and $\alpha=4$.]
	{
		\centering\includegraphics[height=5.9cm]{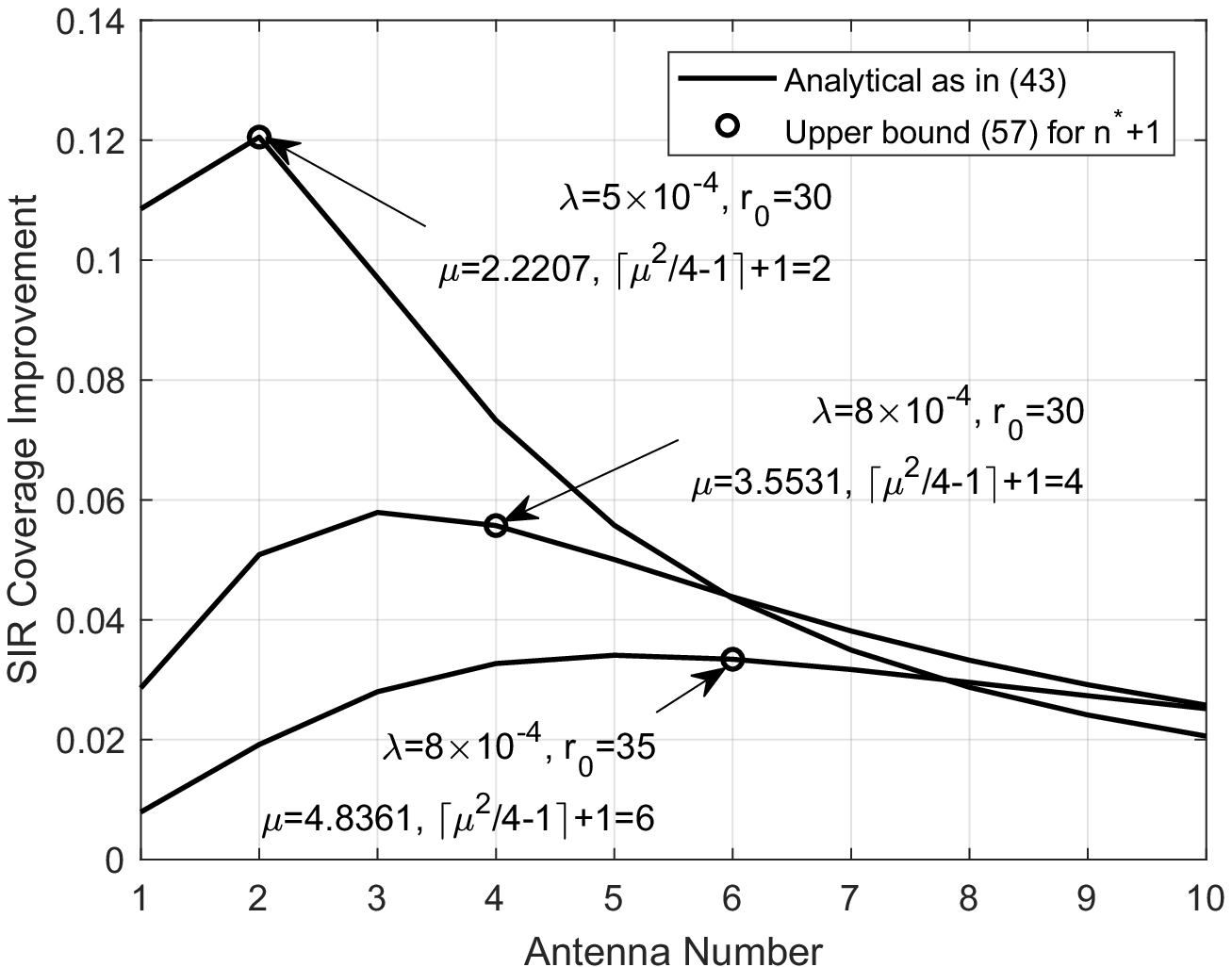}\label{fig6}
	}
	\subfigure[The increment of the SIR coverage probability \eqref{eq50} of ad hoc networks when $\mathrm{SNR}=0$ dB, $\lambda=5\times10^{-3}$, $\theta=\kappa=\beta=1$, and $\alpha=3$.]
	{
		\centering\includegraphics[height=5.9cm]{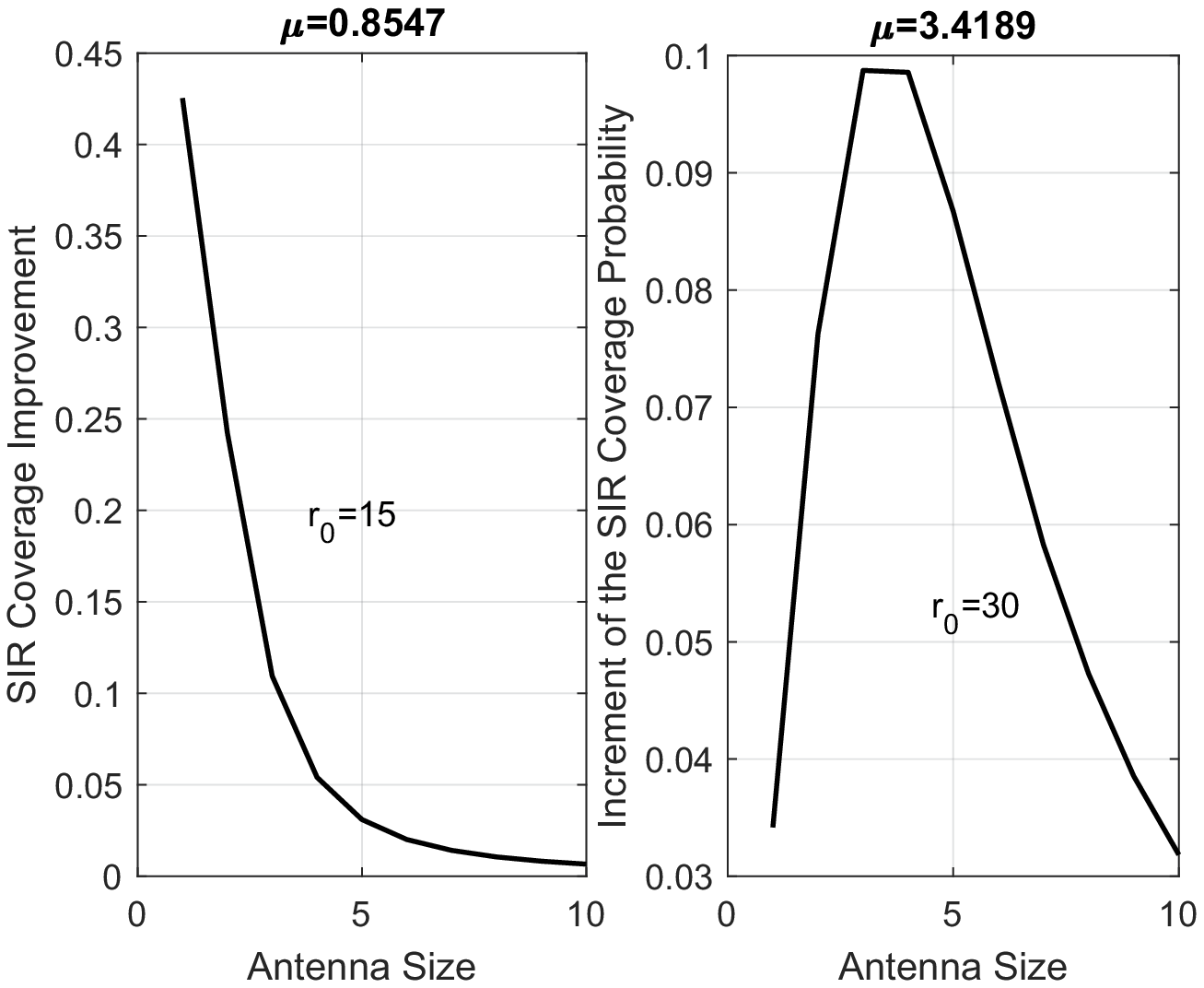}
	\label{fig5}}
\caption{The coverage improvement  in ad hoc networks. When $1-\mu\delta>0$, the coverage improvement monotonically deceases with the antenna index, while there exists a peak value of the coverage improvement when $1-\mu\delta\le0$.}
\end{figure}

The SIR coverage improvement of ad hoc networks when $\alpha=4$ is presented in Fig. \ref{fig6}. The situations when the coverage improvement has a peak value, i.e., $\mu>2$, are of particular interest. It can be discovered that the denser the network (or, equivalently, the longer the dipole distance), the larger the index of the antenna that provides the maximum coverage improvement. Note that we exploit an upper bound in \eqref{eq74}, and therefore $\left\lceil\frac{\mu^2}{4}-1\right\rceil+1$ is an upper bound for the antenna index $n^\star+1$ with the most significant contribution in terms of the coverage improvement. In Fig. \ref{fig6}, we see that this upper bound is very tight, which demonstrates the effectiveness of the result in Proposition \ref{prop5} and the proposed analytical framework.

For the general case $\alpha>2$, although it is difficult to obtain similar analytical results as Proposition \ref{prop5} on the monotonicity of the coverage improvement, a closed-form expression for the coverage improvement is given in the following proposition, which can be used to numerically test the monotonicity property.

\begin{prop}\label{prop4}
	The SIR coverage improvement of the $n+1$-th antenna in ad hoc networks is given by
	\begin{equation}\label{eq50}
	\bar{p}_n=\frac{(-1)^ne^{-\mu}}{n!}\sum_{k=1}^n\rho(n,k)T_k(-\mu)\delta^k,
	\end{equation}
	where $\rho(n,k)$ are the Stirling numbers of the first kind, and $T_k(x)$ denotes the Touchard polynomial \cite{zwillinger2014table}.
\end{prop}
\begin{IEEEproof}
	See Appendix \ref{appE}.
\end{IEEEproof}

\emph{Remark 15:} The Touchard polynomial of order $n$ is obtained when calculating the $n$-th moment of a Poisson distributed random variable. The appearance of such polynomial in \eqref{eq50} is related to the falling factorial and the Taylor expansion of the exponential function.

Fig. \ref{fig5} plots the SIR coverage improvement as the number of antennas increases. It is numerically found that the coverage improvement has two totally different behaviors when the number of antennas increases: 1) When $p_0>p_1$, i.e., $1-\mu\delta>0$, the coverage improvement is monotonically decreasing with the antenna size, which is similar to that in cellular networks. In other words, the coverage improvement would never increase once it decreases at the beginning; 2) When $1-\mu\delta\le0$, the coverage improvement has a peak value $p_{n^\star}$ when the antenna size is enlarged, and the optimal value $n^\star$ can be numerically determined by the closed forms in Proposition \ref{prop4}. This means that adding the $(n^\star+1)$-th antenna is the most effective in terms of the coverage improvement. Note that, for the special case that $\alpha=4$, we have derived in Proposition \ref{prop5} that the coverage improvement monotonically decreases when $\mu<2$. This is a special case of the condition $1-\mu\delta>0$ when $\delta=\frac{1}{2}$, which verifies both the effectiveness of the analytical results in Proposition \ref{prop5} and the reasoning of the conclusion drawn from the simulations results for general cases.

\section{Summary}\label{V}
This paper proposed a unified analytical framework for coverage analysis of  multi-antenna networks. Various tractable analytical results for the coverage probability were demonstrated. In particular, expressions for a general network model was firstly derived. Two typical network models, i.e., cellular and ad hoc networks, were then investigated to demonstrate the generality and effectiveness of the proposed framework. More importantly, system insights, i.e., the impacts of the transmitter density and the antenna size, were analytically revealed via the proposed framework in different multi-antenna networks. Overall, this paper provides a powerful toolbox for the evaluation and design of various  multi-antenna wireless networks, which shall find ample applications.

\appendices
\section{}\label{appB}
Since $g\sim\mathrm{Gamma}(\kappa,\beta)$, i.e., $f_g(u)=\frac{u^{\kappa-1}e^{-\frac{u}{\beta}}}{\beta^\kappa\Gamma(\kappa)}$, according to \eqref{eq26}, \eqref{eq27}, and \eqref{eq28}, we have
\allowdisplaybreaks 
\begin{align}\label{eq48}
c_n&=\frac{(-s)^n}{n!}\frac{\mathrm{d}^n}{\mathrm{d}s^n}\left[1-\frac{\eta(s)}{\pi\lambda r_0^2}\right]=
-\frac{(-s)^n}{n!}\frac{\mathrm{d}^n}{\mathrm{d}s^n}\left\{\delta(sr_0^{-\alpha})^\delta\mathbb{E}_g\left[g^\delta\gamma(-\delta,sr_0^{-\alpha}g)\right]\right\}\nonumber\\
&\overset{(d)}{=}-\frac{(-s)^n}{n!}\frac{\mathrm{d}^n}{\mathrm{d}s^n}\int_1^\infty\mathbb{E}_g\left[\exp\left(-sr_0^{-\alpha}v^{-\frac{\alpha}{2}}g\right)\right]\mathrm{d}v\nonumber\\
&=
-\frac{(-s)^n}{n!}\int_1^\infty\left[\frac{\mathrm{d}^n}{s^n}\frac{1}{\left(1+\beta r_0^{-\alpha}v^{-\frac{\alpha}{2}}s\right)^\kappa}\right]\mathrm{d}v\\
&=
-\frac{\Gamma(\kappa+n)}{\Gamma(\kappa)\Gamma(n+1)}\left(\frac{\tau\beta}{\theta}\right)^\frac{2}{\alpha}\int_{\left(\frac{\tau\beta}{\theta}\right)^{-\frac{2}{\alpha}}}^\infty\frac{\left(v^{-\frac{\alpha}{2}}\right)^n}{\left(1+v^{-\frac{\alpha}{2}}\right)^{\kappa+n}}\mathrm{d}v\nonumber\\
&=\frac{\Gamma(\kappa+n)}{\Gamma(\kappa)\Gamma(n+1)}\frac{\delta}{\delta- n}\left(\frac{\tau\beta}{\theta}\right)^n{}_2F_1\left(n+\kappa,n-\delta;n+1-\delta;-\frac{\tau \beta}{\theta}\right),\nonumber
\end{align}
where {\color{black} (d) follows from the definition of the lower incomplete Gamma function $\gamma(s,x)$, and} the last equality follows from the integral representation of the hypergeometric function \cite[Sec. 9.14]{zwillinger2014table}, which completes the proof.

\section{}\label{appA}
According to the two steps of applying Theorem \ref{th1} presented in Remark 4, first we calculate the conditional Laplace transform, expressed as
\begin{equation}
\mathcal{L}(s)=\exp\left\{-2\pi\lambda\int_0^\infty{\left(1-\mathbb{E}_{g}[\exp(-sgv^{-\alpha})]\right)}v\mathrm{d}v\right\}.
\end{equation}
To obtain a coverage probability expression for arbitrarily distributed interferers' power gains, we propose to swap the order of the integral and the expectation. In this way, part of the exponent is given by
\begin{equation}
\begin{split}
&\phantom{=\,\,\,}2\mathbb{E}_{g}\left\{\int_0^\infty{\left[1-\exp(-sgv^{-\alpha})\right]}v\mathrm{d}v\right\}\\
&=\mathbb{E}_g\left\{(sg)^\frac{2}{\alpha}\frac{2}{\alpha}\int_0^1\frac{v}{1-v}\left[-\ln(1-v)\right]^{-\frac{2}{\alpha}-1}\mathrm{d}v\right\}=\mathbb{E}_g\left\{(sg)^\delta\Gamma(1-\delta)\right\}.
\end{split}
\end{equation}
Therefore, the log-Laplace transform  can be written as
\begin{equation}\label{eq47}
\eta(s)=-\pi\lambda\Gamma(1-\delta)s^\delta\mathbb{E}_g\left[g^\delta\right],
\end{equation}
and the non-zero entries of $\mathbf{A}_M$ are determined by
\begin{equation}\label{eq56}
a_n=\frac{(-s)^n}{n!}\eta^{(n)}(s)=-\frac{(-1)^n}{n!}
\pi\lambda r_0^2\Gamma(1-\delta)(\delta)_n\left(\frac{\tau}{\theta}\right)^\delta\mathbb{E}_g\left[g^\delta\right].
\end{equation}
Since there is no need to take an expectation over $r_0$ in the ad hoc network model, the derivation steps similar to \eqref{eq31} and \eqref{eq36} are unnecessary, and the proof is complete.

\section{}\label{appC}
According to \eqref{eq37}, the Laplace transform of noise and interference is
\begin{equation}
\mathcal{L}(s)=p_0=e^{\eta(s)}=\exp\left(-s\sigma_\mathrm{n}^2-\pi\lambda\Gamma(1-\delta)s^\delta\mathbb{E}_g\left[g^\delta\right]\right).
\end{equation}
Note that $\Gamma(1-\delta)$ is a positive term due to the fact that $0<\delta<1$. Hence, the Laplace transform $p_0$ is a convex and monotonically decreasing function with respect to the transmitter density $\lambda$. 

Furthermore, according to \eqref{an3}, the signs of $\{a_n\}_{n=1}^{M-1}$ are critical, i.e.,
\begin{equation}\label{eq61}
a_n=
-\frac{(-1)^n}{n!}(\delta)_n\pi\lambda \Gamma(1-\delta)s^\delta\mathbb{E}_g\left[g^\delta\right]+s\sigma_\mathrm{n}^2\mathds{1}(n=1).
\end{equation}
Since $(-1)^n(\delta)_n=(-\delta)^{(n)}<0$ with $(x)^{(n)}$ denoting the rising factorial, we have $a_n>0$ for $1\le n\le M$.
Recall that the recursive relations between $\{p_n\}_{n=1}^{M-1}$ are
\begin{equation}
p_n=\sum_{i=0}^{n-1}\frac{n-i}{n}a_{n-i}p_i.
\end{equation}
Since the term $\frac{n-i}{n}a_{n-i}$ are positive, it turns out that all $\{p_n\}_{n=1}^{M-1}$ have the same monotonicity and convexity with respect to $\lambda$. Recalling that $p_\mathrm{c}(\tau)=\sum_{n=0}^{M-1}p_n$, the monotonicity and concavity in Corollary \ref{coro3} has been proved. Next, we prove the expression \eqref{eq41}. 

We first write $\mathbf{A}_M^\prime$ in the form
\begin{equation}
\mathbf{A}_M^\prime=a_0^\prime\mathbf{I}_M+(\mathbf{A}_M^\prime-a_0^\prime\mathbf{I}_M).
\end{equation}
Since $\mathbf{A}_M^\prime$ is a lower triangular Toeplitz matrix, the second part is a nilpotent matrix, i.e., $(\mathbf{A}_M^\prime-a_0^\prime\mathbf{I}_M)^n=\mathbf{0}$ for $n\ge M$. Hence, according to the properties of matrix exponential, we have
\begin{equation}\label{eq62}
e^{\mathbf{A}_M}=e^{\lambda\mathbf{A}_M^\prime}=e^{a_0^\prime\lambda}\cdot
\sum_{n=0}^{M-1}\frac{1}{n!}\left[\lambda\left(\mathbf{A}_M^\prime-a_0^\prime\mathbf{I}_M\right)\right]^n.
\end{equation}
Since it has been shown that $a_n^\prime>0$ for $n\ge1$, $\mathbf{A}_M^\prime-a_0^\prime\mathbf{I}_M$ is a strictly lower triangular Toeplitz matrix with all positive entries, and so are the matrices $(\mathbf{A}_M^\prime-a_0^\prime\mathbf{I}_M)^n$. Therefore,
\begin{equation}
\left\Vert e^{\lambda\mathbf{A}_M^\prime}\right\Vert_1=
e^{a_0^\prime\lambda}\cdot
\sum_{n=0}^{M-1}\frac{1}{n!}\left[\lambda^n\left\Vert\left(\mathbf{A}_M^\prime-a_0^\prime\mathbf{I}_M\right)^n\right\Vert_1\right],
\end{equation}
which completes the proof of Corollary \ref{coro3}.


\section{}\label{appD}
According to Theorem \ref{the1}, the outage probability is $p_\mathrm{o}(\tau)=1-\sum_{n=0}^{M-1}\bar{p}_n$, then
\begin{equation}
\underset{M\to\infty}{\lim}\frac{p_\mathrm{o}(M+1)}{p_\mathrm{o}(M)}=1-\underset{M\to\infty}{\lim}\frac{\bar p_M}{1-\sum_{n=0}^{M-1}\bar{p}_n}=1-\underset{M\to\infty}{\lim}\frac{1}{1-\sum_{n=M}^\infty\frac{\bar{p}_n}{\bar p_M}}.
\end{equation}
Since $r_\mathrm{c}$ is the radius of convergence of the power series $\bar{P}(z)$, i.e., $r_\mathrm{c}=\underset{n\to\infty}{\lim}\frac{\bar{p}_n}{\bar p_{n+1}}$, the above equation can be further simplified as
\begin{equation}
\underset{M\to\infty}{\lim}\frac{p_\mathrm{o}(M+1)}{p_\mathrm{o}(M)}=1-\underset{M\to\infty}{\lim}\frac{1}{\sum_{n=0}^\infty\left(\frac{1}{r_\mathrm{c}}\right)^n}=\frac{1}{r_\mathrm{c}}.
\end{equation}
According to \eqref{eq48}, the coefficients in the power series $C(z)$ are given by
\begin{equation}
c_n=\frac{(-s)^n}{n!}c_0^{(n)}(s),
\end{equation}
where $c_0(s)=-\delta(sr_0^{-\alpha})^\delta \mathbb{E}_g\left[g^\delta\gamma(-\delta,sr_0^{-\alpha}g)\right]$. By reversely applying the Taylor expansion, the power series $C(z)$ can be written as
\begin{equation}
C(z)=\sum_{n=0}^\infty c_n z^n=\sum_{n=0}^\infty\frac{(-sz)^n}{n!}c_0^{(n)}(s)=c_0((1-z)s).
\end{equation}
Recalling that in \eqref{eq31} we proved that $\bar{P}(z)=\frac{1}{C(z)}$, thus the radius of convergence of $\bar{P}(z)$ is the solution of the equation $C(r_\mathrm{c})=c_0((1-r_\mathrm{c})s)=0$, which is equivalent to \eqref{equation47}.

Next, we prove that the solution $r_\mathrm{c}$ to equation \eqref{equation47} is larger than 1. 
The left hand side of \eqref{equation47} can be rewritten as
\begin{equation}\label{eq75}
\mathbb{E}_g\left[{}_1F_1\left(-\delta;1-\delta;\frac{(r_\mathrm{c}-1)\tau}{\theta}g\right)\right]=1+\delta\mathbb{E}_g\left[\int_0^1\frac{1-e^{\frac{(r_\mathrm{c}-1)\tau}{\theta}gv}}{v^{1+\delta}}\mathrm{d}v\right].
\end{equation}
Since $0<\delta<1$, $\tau>0$, $\beta>0$, and $g$ is assumed as a non-negative random variable with arbitrary distributions, it is seen from \eqref{eq75} that $C(r_\mathrm{c})$ is a monotonically decreasing function of $r_\mathrm{c}$. Furthermore, it is easy to check that, when $r_\mathrm{c}=1$, we have $C(1)=1$. Following the monotonicity of $C(r_\mathrm{c})$ and the fact that $C(1)>0$, we conclude that there exists only one solution of \eqref{equation47} that is larger than 1.

\section{}\label{appF}
According to \eqref{eq56}, we have
\begin{equation}
\begin{split}
A(z)&=\sum_{n=0}^\infty a_n z^n
=\sum_{n=0}^\infty\frac{(-sz)^n}{n!}\eta^{(n)}(s)=\eta((1-z)s)\\
&=-\pi\lambda r_0^2\Gamma(1-\delta)\mathbb{E}_g\left[g^\delta\right](1-z)^\delta=-\mu(1-z)^\delta.
\end{split}
\end{equation}
Therefore, with the formulas \eqref{eq33} and \eqref{eq31}, we have the closed-form expression
\begin{equation}\label{eq79}
\bar{P}(z)=P(z)=e^{{A(z)}}=e^{-\mu(1-z)^\delta}.
\end{equation}
When $\alpha=4$, i.e., $\delta=1/2$, the power series $\bar{P}(z)$ is given by
$\bar{P}(z)=\sum_{n=0}^\infty \bar{p}_nz^n=e^{-\mu\sqrt{1-z}}$.
According to the definition of the modified Bessel function of the second kind $K_n(x)$ \cite[pp. 39]{abramowitz1966handbook}, we have
\begin{equation}
\bar{p}_n=\sqrt{\frac{2\mu}{\pi}}\frac{(\mu/2)^n}{n!}K_{n-\frac{1}{2}}(\mu).
\end{equation}
Then, define the ratio to test the monotonicity as
\begin{equation}\label{eq74}
\frac{\bar p_{n+1}}{\bar{p}_n}=\frac{\mu}{2(n+1)}\frac{K_{n+\frac{1}{2}}(\mu)}{K_{n-\frac{1}{2}}(\mu)}\overset{(e)}{\le}\frac{n+\sqrt{n^2+\mu^2}}{2(n+1)},
\end{equation}
where the inequality adopted in (e) comes from \cite[Th. 1]{segura2011bounds}. Finally, it can be checked that $\frac{n+\sqrt{n^2+\mu^2}}{2(n+1)}<1$ when $n>\frac{\mu^2}{4}-1$, which completes the proof.

\section{}\label{appE}
By performing coefficient extraction to \eqref{eq79},
\begin{equation}
\begin{split}
\bar{P}(z)=e^{\mu(1-z)^\delta}=\sum_{k=0}^\infty\frac{\mu^k(1-z)^{\delta k}}{k!}=\sum_{k=0}^\infty\frac{\mu^k}{k!}\sum_{n=0}^\infty\frac{(-1)^n}{n!}(\delta k)_nz^n=
\sum_{n=0}^\infty\left[\frac{(-1)^n}{n!}\sum_{k=0}^\infty \frac{\mu^k}{k!}(\delta k)_n\right]z^n,
\end{split}
\end{equation}
we have
\begin{equation}
\begin{split}
\bar{p}_n&=\frac{(-1)^n}{n!}\sum_{k=0}^\infty \frac{(-\mu)^k}{k!}(\delta k)_n\overset{(f)}{=}\frac{(-1)^n}{n!}\sum_{k=0}^\infty \frac{(-\mu)^k}{k!} \sum_{p=0}^n\rho(n,p)(\delta k)^p\\
&=\frac{(-1)^n}{n!}\sum_{p=0}^n\rho(n,p)\delta^p\sum_{k=0}^\infty \frac{(-\mu)^k}{k!}  k^p\overset{(g)}{=}\frac{(-1)^ne^{-\mu}}{n!}\sum_{k=1}^n\rho(n,k)T_k(-\mu)\delta^k,
\end{split}
\end{equation}
where steps (f) and (g) reversely apply the definition of the Stirling numbers of the first kind and the Touchard polynomial, respectively.

\bibliographystyle{IEEEtran}
\bibliography{bare_jrnl}
\end{document}